\documentclass[aip,jcp,reprint,longbibliography]{revtex4-1} 
\usepackage{graphicx}
\usepackage{enumitem}
\usepackage{mathtools}
\usepackage{graphicx}   
\usepackage{color}
\usepackage{xcolor}
\usepackage{amssymb}

\usepackage{tabularx}

\usepackage[normalem]{ulem}
\usepackage{caption}
\usepackage{subcaption}
\usepackage{comment}
\usepackage{bm}
\usepackage{array}
\usepackage{booktabs}
\usepackage{multirow}

\usepackage{amsmath}

\usepackage{booktabs}

\usepackage[breaklinks=true]{hyperref}
\usepackage{breakcites}

\begin{document}
\title{Exponentially Tilted Thermodynamic Maps (expTM): Predicting Phase Transitions Across Temperature, Pressure, and Chemical Potential}
	
\author{Suemin Lee*}
 \affiliation{Biophysics Program and Institute for Physical Science and Technology,
 University of Maryland, College Park 20742, USA}
   \affiliation{University of Maryland Institute for Health Computing, Bethesda, Maryland 20852, USA}

\author{Ruiyu Wang*}
 \affiliation{Institute for Physical Science and Technology,
 University of Maryland, College Park 20742, USA}
 
\author{Lukas Herron}

\affiliation{Biophysics Program and Institute for Physical Science and Technology,
 University of Maryland, College Park 20742, USA}
   \affiliation{University of Maryland Institute for Health Computing, Bethesda, Maryland 20852, USA}

 \author{Pratyush Tiwary\footnote{Corresponding author.}}
 \email{ptiwary@umd.edu}
\affiliation{Biophysics Program and Institute for Physical Science and Technology,
 University of Maryland, College Park 20742, USA}
  \affiliation{University of Maryland Institute for Health Computing, Bethesda, Maryland 20852, USA}
 \affiliation{Department of Chemistry and Biochemistry and Institute for Physical Science and Technology,
 University of Maryland, College Park 20742, USA}

\date{\today}
	
\begin{abstract}
\textbf{Abstract\newline}
Predicting and characterizing phase transitions is crucial for understanding generic physical phenomena such as crystallization, protein folding and others. However, directly observing phase transitions is not always easy, and often one has limited observations far from the phase boundary and measured under some specific thermodynamic conditions. 
In this study, we propose a statistical physics and Generative AI driven framework that can take such limited information to generate samples of different phases under arbitrary thermodynamic conditions, which we name Exponentially Tilted Thermodynamic Maps (expTM). The central idea is to map collected data into a tractable simple prior  expressed as an exponentially tilted Gaussian. We demonstrate how the variance and mean of the prior can be correlated with pairs of thermodynamic control variables, including temperature, pressure, and chemical potential. This gives us the ability to generate thermodynamically correct samples under any values of the control variables. 
To demonstrate the practical applicability of this approach, we use expTM to sample the lattice gas models with the Grand Canonical ensemble, capturing phase transitions under varying chemical potentials and temperatures. We further demonstrate how expTM can model the isothermal-isobaric ensemble, with which we predict different phases of CO$_2$ under varying pressure conditions. Both examples are trained on very limited data far from the phase boundary. These results establish expTM as a robust tool for understanding phase transitions across diverse thermodynamic conditions requiring only a small number of observations.

\end{abstract}

\maketitle
	
\section{Introduction}
\label{sec:introduction}

Phase transitions are ubiquitous in nature, with the transformation of ice melting into water or water evaporating serving as classic examples.~\cite{stanley1987introduction} From a materials science perspective, understanding the atomic level change of the systems is essential not only for discovering novel materials but also for understanding protein folding and conformational changes.\cite{dill2012the}
To tackle these important problems, common computational methods are Monte Carlo (MC) and Molecular Dynamics (MD) simulations, alongside experimental techniques.~\cite{James1980Monte,Karplus2002molecular,sosso2016crystal,beyerle2023recent,wang2024atomic} However, these traditional approaches face limitations, such as time-scale constraints and limited data often making it difficult to capture every aspect of a system. Recently, the rapid development of generative artificial intelligence (AI) has gained interest in using new data-driven strategies to advance our understanding in various fields, including phase transitions.

Among various generative AI frameworks, diffusion models—originally inspired by non-equilibrium thermodynamics—have gained particular interest.\cite{dickstein2015deep,ho2020denoising,song2021scorebased,song2021denoising} Sohl-Dickstein et al.\cite{dickstein2015deep} initially proposed the model to generate synthetic data by reversing a non-equilibrium diffusion process, laying the groundwork for subsequent refinements.\cite{ho2020denoising,song2021scorebased,song2021denoising} In brief, given data samples from a hard-to-sample {\it target} distribution, a diffusion model parameterizes a stochastic process that transports samples from an easy-to-sample {\it prior} distribution to the {\it target} distribution with a neural network. Initially, this idea was used to generate images from random noise, but it has more recently seen broad application across domains, including within the molecular sciences: GeoDiff\cite{xu2022geodiff} and GeoLDM\cite{xu2023geometric} generate molecular conformers, DiffDock\cite{corso2023diffdock} and NeuralPlexer\cite{qiao2024state} predict binding poses of ligands with proteins, and AlphaFold3\cite{Abramson2024accurate} predicts general biomolecular structures. These successes highlight the potential of diffusion models to address scientific challenges.\cite{tiwary2024generative}

Building on this foundation, our previous work introduced Thermodynamic Maps (TM),\cite{herron2024inferring} a framework that builds upon diffusion models to characterize phase transitions and accurately identify temperature-dependent properties, even with minimal data from stable phases. TM does this by implicitly encoding temperature-dependent fluctuations through the canonical ensemble. TM has successfully been demonstrated its effectiveness in identifying phase transitions across diverse systems, including the Ising model, RNA conformations, and the Gay-Berne model \cite{herron2024inferring, beyerle2024inferring}. Given that TM has been shown to be capable of generating observations extrapolating across temperatures, it is natural to wonder how far, if at all, can the procedure be generalized to treat extrapolation in the space of other important environmental variables—such as pressure and chemical potential that can govern a system’s behavior including phase transitions.

By offering a menu of ensembles, statistical mechanics provides the language and machinery for expressing how different environmental variables can fluctuate in tandem. Specifically, the Grand Canonical ensemble allows for fluctuations in particle number and is applicable when a system exchanges particles with its surroundings with the chemical potential as a control variable. Similarly, in the isothermal-isobaric ensemble both temperature and pressure are held constant and the volume of the system can fluctuate. 

To develop a Generative AI framework that can extrapolate physics across arbitrary thermodynamic variables including temperature, pressure and chemical potential, here, we extend the TM framework proposed in Ref. \onlinecite{herron2024inferring}. This is achieved by incorporating an exponential tilting factor into the Gaussian prior distribution used in  Ref. \onlinecite{herron2024inferring}. This modification, which we term Exponentially Tilted Thermodynamic Maps (expTM), enables simultaneous control over two distinct environmental parameters and broadens the applicability to a range of different thermodynamic ensembles. In this paper, we present the theoretical foundations of expTM and demonstrate its effectiveness through two key applications. First, we apply expTM to a lattice gas model within the Grand Canonical ensemble, where it accurately reproduces the density of particles and critical behavior. Second, we employ expTM to predict pressure-induced phase transitions in CO$_2$ under the isothermal-isobaric ensemble, successfully delineating phase boundaries and capturing intermediate states with minimal training data.

The remainder of the paper is organized as follows: In Section Sec.\ref{sec:Methods}, we review diffusion models, exponential tilting, the original TM framework and introduce theoretical framework for thermodynamic maps generalized via exponential tilting. In Sec.\ref{sec:Results}, we demonstrate two applications of expTM by inferring phase transitions of a lattice gas (under the Grand Canonical ensemble) and CO$_2$ (under the isothermal-isobaric ensemble) from minimal training data. Finally, in Sec.~\ref{sec:conclusion}, we summarize our results and discuss future directions.

\section{Methods}
\label{sec:Methods}

In this section, we first review diffusion models (Sec. \ref{sec:diffusion_models}), followed by the effect of exponential tilting on diffusion models (Sec.\ref{sec:Diffusion_models_exponentially_tilting}). In Sec. \ref{sec:ThermodynamicMaps}, we recapitulate the thermodynamic maps (TM) approach. In Sec.\ref{sec:expTM}, we further propose expTM via exponential tilting and discuss how expTM can be generalized to different ensemble systems, such as isothermal-isobaric ensembles and Grand Canonical ensembles, respectively in Sec.\ref{sec:expTM_NPT}, and Sec.\ref{sec:expTM_muVT}.

\subsection{Diffusion models}
\label{sec:diffusion_models}

Diffusion models have emerged as a powerful generative framework for modeling complex data distributions.~\cite{dickstein2015deep,song2019generative,ho2020denoising,song2021scorebased} The central idea underlying diffusion models is to define a forward-time diffusion process that adds noise to samples from the target distribution $\mathcal{D}$, gradually transforming them into samples from a tailored prior distribution which is easy to sample. A neural network learns to approximate the reverse diffusion process, gradually denoising samples to recover the original distribution. This enables generating further samples from the target distribution by first sampling from the prior and then simulating the time-reversed process.

The forward-time diffusion is conventionally an Ornstein-Uhlenbeck\cite{uhlenbeck1930theory,Doob1942Brownian} process modeled as a stochastic differential equation (SDE) of the form
\begin{equation} 
\mathrm{d}\mathbf{x}_t = \underbrace{-g(t) \mathbf{x}_t \mathrm{d}t}_{\mathrm{drift}} + \underbrace{\sqrt{2g(t)} \mathrm{d}\mathbf{B}_t}_{\mathrm{noise}}.
\label{eq:diffusion_forward_process}
\end{equation}
The state is denoted by $\mathbf{x}\in \mathbb{R}^d$ and is implicitly a function of the time $t$ starting from 0 and going to 1. The term $g(t)$ is the noise schedule governing the rate of convergence, and is typically chosen so that the process is sufficiently converged to the desired prior at $t=1$. The drift term controls the mean-reverting behavior of the process, drawing $\mathbf{x}$ towards zero on average. Stochastic fluctuations $\mathrm{d}\mathbf{B}_t$ are modeled as standard Brownian motion. Together in Eq.~\ref{eq:diffusion_forward_process} they define a stochastic process that relaxes to $\mathcal{N}(0,I_d)$ at $t=1$ for any initial condition at $t=0$.

With the forward process defined, one can show that a time-reverse process exists~\cite{anderson1982reverse} and is an SDE of the form
\begin{equation}
\begin{aligned}
\mathrm{d}\mathbf{x}_\tau = -g(\tau)\left[\mathbf{x}_\tau + \nabla \log p_\tau
(\mathbf{x})\right] \mathrm{d}\tau + \sqrt{2g(\tau)} \mathrm{d}\mathbf{B}_\tau.
\label{eq:diffusion_reverse_process}
\end{aligned}
\end{equation}

Here, $p_\tau(\mathbf{x})$ denotes the marginal probability density of $\mathbf{x}$ at reverse time $\tau \equiv 1-t$. Its spatial gradient $\nabla \log p_\tau(\mathbf{x})$ is the \emph{score} -- the central quantity of interest, and is estimated from realizations of the forward-time process using a neural network. If the score estimate is sufficiently accurate, simulating the reverse-time process starting from initial conditions distributed according to $\mathcal{N}(0, I_d)$ yields endpoints sampled from the target distribution~\cite{hyvarinen2005Estimation,vincent2011connectionv, song2019generative}.

\begin{figure*}%[htbp]
\captionsetup{skip=6pt}
\includegraphics[width=.87\textwidth]{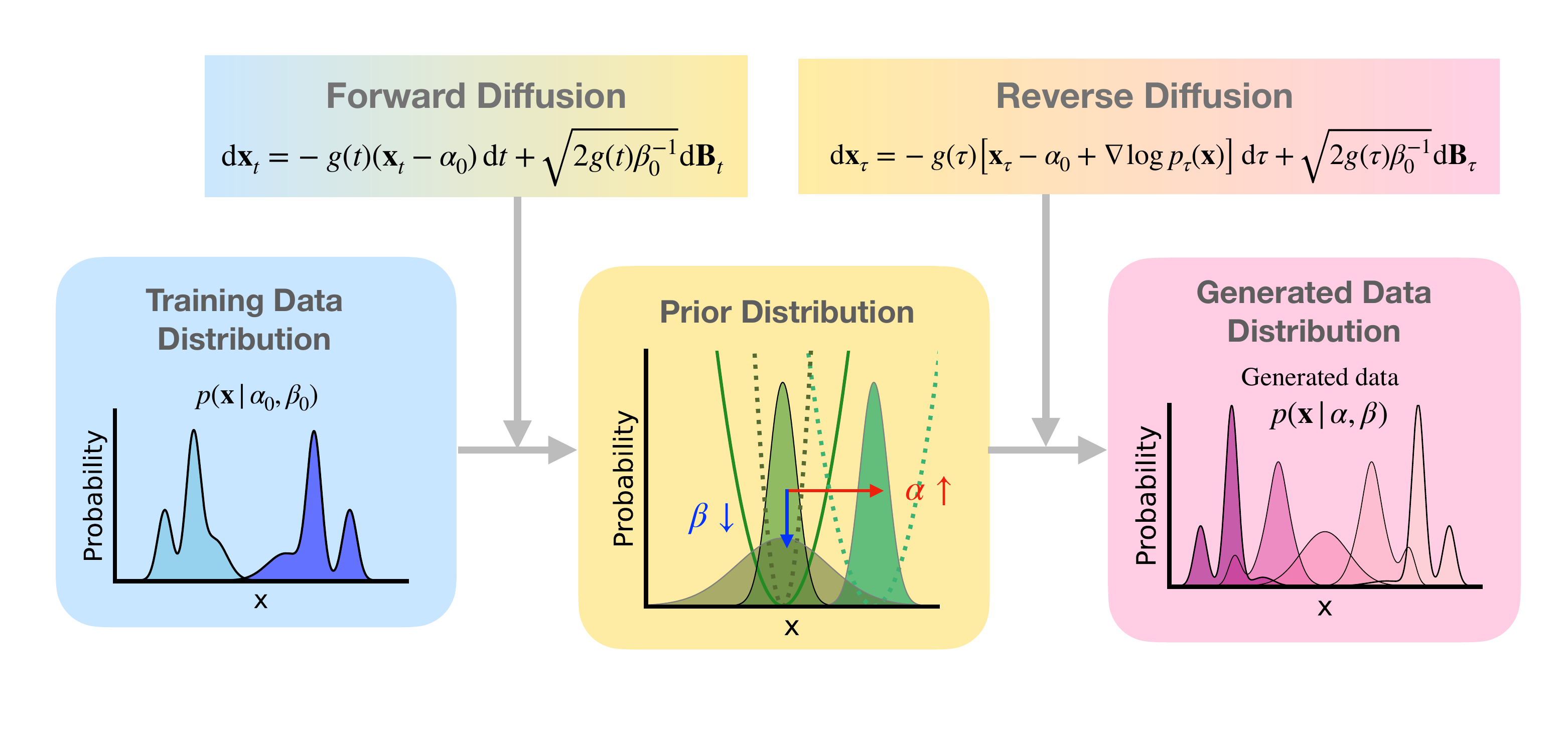}
    \caption{Exponentially Tilted Thermodynamic Maps (expTM) Schematic: The diagram illustrates how training data from known thermodynamic conditions, $(\alpha_0,\beta_0)$ (blue box), are mapped to a tilted Gaussian prior (yellow box) via a forward diffusion, and then mapped back from the prior to produce new samples $p(\mathbf{x}\mid \alpha,\beta)$ (pink box) under any desired thermodynamic parameters $(\alpha,\beta)$. Here, $\alpha$ represents the thermodynamic variable of interest (e.g., pressure $P$ or chemical potential $\mu$), and $\beta$ denotes the inverse temperature. In the prior, each sample $\mathbf{x}$ follows $\mathcal{N}(\alpha,\beta^{-1}{I}_d)$, thereby reflecting both a mean shift $\alpha$ and a variance proportional to $\beta^{-1}$.  By learning the forward and reverse mappings, expTM can produce samples $p(\mathbf{x}\!\mid\!\alpha,\beta)$ over a wide range of parameter values of interest.}
    \label{fig:Model architecture} 
\end{figure*}

\subsection{Exponentially tilted diffusion models}
\label{sec:Diffusion_models_exponentially_tilting}
A typical choice of prior distribution for generative models is Gaussian distribution for its simplicity and tractability. However, in certain cases, the Gaussian prior is not expressive enough to capture features from the training data that lead to poor representations of the latent space.~\cite{Hoffman2016Elbo}
To address the limitation, alternative prior distributions have been proposed to improve the accuracy and efficiency of the generative models. \cite{Nam2025Flow,lee2022priorgrad,jing2023eigenfol,Guan2023decompdiff,xiao2020exponential}. In particular, Ref.~\onlinecite{floto2023the} uses tilted Gaussian distributions in variational autoencoder models to enhance out-of-distribution detection of latent representations. In the same vein as these other works, here, in order to improve expressiveness, we extend our prior distribution by integrating a modified formulation of exponential tilting into the diffusion prior.

Exponential tilting modifies a base probability distribution $p(\mathbf{x})$ by introducing an exponential factor that shifts its mean.\cite{Fuh2024efficient} For instance, consider a standard normal distribution:

\begin{equation}
    p(\mathbf{x}) = \frac{1}{\sqrt{2\pi}} e^{-\frac{\mathbf{x}^2}{2}}.
    \label{eq:standard_normal_distribution}
\end{equation}

\noindent which we denoted as $\mathcal{N}(0,\mathcal{I}_d)$. We multiply Eq.~\ref{eq:standard_normal_distribution} by the exponential factor $e^{\alpha \mathbf{x}}$, where the exponential part is composed of a simple linear term $\alpha \mathbf{x}$, written in the following form: 

\begin{equation}
    p(\mathbf{x}) = \frac{1}{\sqrt{2\pi}} e^{-\frac{\mathbf{x}^2}{2}}e^{\alpha \mathbf{x}}.
    \label{eq:exponetial_tilted_distribution}
\end{equation}

\noindent After the rearrangement of the Eq.~\ref{eq:exponetial_tilted_distribution}, this causes the distribution $p(\mathbf{x})$ to be tilted, where the mean shifts to the new distribution $\mathcal{N}(\alpha,\mathcal{I}_d)$.

In fact, Gaussian exponential tilting is closely related to stochastic processes. The shift in mean $\alpha$ corresponds to introducing an additional drift term into the underlying OU process. In other words, introducing an additional drift term $\alpha$ can be viewed as an external force acting on the potential. Consequently, applying an exponential tilting factor, i.e. $e^{\alpha x}$,  to a Gaussian prior can be interpreted as embedding an adjustable drift into the diffusion process, leading to convergence toward a new mean. This effect can be incorporated into the previous Eq.~\ref{eq:diffusion_forward_process} modified in the following form:

\begin{equation} 
\mathrm{d}\mathbf{x}_t = -g(t) (\mathbf{x}_t -\alpha)\,\mathrm{d}t + \sqrt{2g(t)} \mathrm{d}\mathbf{B}_t, \quad \mathbf{x}_0 = x.
\label{eq:forward_force}
\end{equation}

The inclusion of exponential prefactor changes the mean from zero to $\alpha$, i.e. $\langle \mathbf{x}_t \rangle = \alpha$, indicating the mean converges to $\alpha$. The variance of $g(t)$ remains the same, as it is independent of the mean. Similarly, the tilted reverse process is expressed as:

\begin{equation}
\begin{aligned}
\mathrm{d}\mathbf{x}_\tau &= -g(\tau)\left[\mathbf{x}_\tau -\alpha + \nabla \log p_\tau
(\mathbf{x})\right] \mathrm{d}\tau + \sqrt{2g(\tau)} \mathrm{d}\mathbf{B}_\tau, 
\\
\quad &\mathbf{x}_{\tau=0}  \sim \mathcal{N}(\alpha, \mathcal{I}_d)
\label{eq:reverse_force}
\end{aligned}
\end{equation}

Here, the Eq.~\ref{eq:reverse_force} mirrors the structure of the original reverse process of Eq.~\ref{eq:diffusion_reverse_process}, with the additional drift term $\alpha$ accounting for the shift in the mean of the distribution. This indicates that during the reverse process, the samples are drawn from a normal distribution with mean $\alpha$ and covariance $I_d$. The inclusion of additional exponentially tilted prefactors enhances the expressiveness of the diffusion model, increasing diversity in the generative process compared to the original Gaussian prior.

\subsection{Thermodynamic maps (TM)}
\label{sec:ThermodynamicMaps}
In this section, we briefly review the theoretical aspects of the original Thermodynamic Maps (TM) and further discuss how the TM effectively captures the canonical ensemble through a correspondingly defined prior distribution. Notably, TM offers two key advantages over other diffusion-based models. First, with a limited amount of data far from the phase boundary, the model can predict first-order transitions in the simple Ising model and extend the predictions to RNA conformations at various temperatures. Second, this predictive capability is supported by a robust theoretical framework. 
Under the diffusion model framework, features are mapped onto a coordinate space $\mathbf{x}$ while their fluctuations across all components are simultaneously mapped onto an abstract temperature space $\boldsymbol{\beta^{-1}}$, which has the same dimensionality as $\mathbf{x}$. To avoid confusion in our notation, we use $x$ to denote a microstate in a particular ensemble, and $\mathbf{x}\in \mathbb{R}^D$ to represent the $D$-dimensional microscopic features of the TM diffusion model.
In the following, we examine the prior distribution in the original TM and show how the mapping to data collected under the canonical ensemble is carried out.

In the canonical (NVT) ensemble, the probability of finding the system in a specific microstate $ x $ is governed by the Boltzmann distribution:
\begin{equation}
    p(x) = \frac{1}{Z} e^{-\beta E(x)},
    \label{eq:nvt_ensemble1}
\end{equation}

\noindent where $ E(x) $ is the energy function, $ \beta = \frac{1}{k_B T} $ is the inverse temperature, and $ Z = \int e^{-\beta E(x)} dx $ is the partition function. For the prior in TM, we assume the energy function to be a simple harmonic potential $ E(x) = \frac{x^2}{2} $, with the prior probability given by:
\begin{equation}
    p(x) = \frac{1}{Z_I} e^{-\frac{\beta}{2} x^2},
    \label{eq:nvt_ensemble2}
\end{equation}

\noindent Here the normalization constant $ Z_I = \int e^{-\frac{\beta}{2} x^2} dx = \sqrt{\frac{2\pi}{\beta}} $. From this, the mean and variance of $ x $ are calculated as $\langle x \rangle = 0$, $\langle x^2 \rangle = \beta^{-1}$. By assuming the energy function to take the form of a simple harmonic potential, the diffusion prior naturally aligns with data collected at different temperatures using the canonical ensemble. In TM\cite{herron2024inferring}, the forward and reverse diffusion processes are applied to feature states $ \mathbf{x} $ rather than  microstate variables $ x $. These forward and reverse diffusion processes respectively are defined as,

\begin{equation}
\begin{pmatrix}
\mathrm{d}\mathbf{x}\\
\mathrm{d}\boldsymbol{\beta^{-1}}
\end{pmatrix}
=
-g(t)
\begin{pmatrix}
\mathbf{x}\\
\boldsymbol{\beta^{-1}}
\end{pmatrix}
\mathrm{d}t
+ \sqrt{2 g(t)} 
\begin{pmatrix}
\sqrt{\boldsymbol{\beta_0^{-1}}} \\
1
\end{pmatrix}
\mathrm{d}\mathbf{B}_t
\label{eq:tm_forward}
\end{equation}

\begin{equation}
\begin{aligned}
\begin{pmatrix}
\mathrm{d}\mathbf{x}\\
\mathrm{d}\boldsymbol{\beta^{-1}}
\end{pmatrix}
&=
-g(\tau) 
\left[
\begin{pmatrix}
\mathbf{x}\\
\boldsymbol{\beta^{-1}}
\end{pmatrix}
+ \mathrm{s}_\theta(\mathbf{x}, \boldsymbol{\beta^{-1}}, \tau)
\right] \mathrm{d}\tau \\
&\quad + \sqrt{2 g(\tau)} 
\begin{pmatrix}
\sqrt{\boldsymbol{\beta_0^{-1}}} \\
\mathbf{1}
\end{pmatrix}
\mathrm{d}\mathbf{B}_\tau.
\end{aligned}
\label{eq:tm_backward}
\end{equation}

Compared to the original score-based model in Eqs.~\ref{eq:diffusion_forward_process} and \ref{eq:diffusion_reverse_process}, which operates solely in the $x$ space, the TM framework defines both the forward and reverse processes in the joint $(\mathbf{x},\boldsymbol{\beta^{-1}})$ space. Here, the vector $\boldsymbol{\beta^{-1}}$ with same dimensionality as $\mathbf{x}$ quantifies component-wise fluctuations in
the feature space $\mathbf{x}$, as a function of the bath temperature. In particular, Eq.~\ref{eq:tm_forward} for the feature state $\mathbf{x}$ corresponds to the forward SDE in Eq.~\ref{eq:diffusion_forward_process}, with variance in its prior determined by the bath temperature $\beta_{0}^{-1}$. This forward process drives the system toward a Gaussian distribution $\mathcal{N}(0,\beta_0^{-1})$, where the variance in the prior reflects the bath temperature at which data was collected. In the reverse process, described by Eq.~\ref{eq:tm_backward}, the score function explicitly depends on the $\beta$ variable, introducing a temperature-dependent correction.  This illustrates that the prior distribution for both the features $\mathbf{x}$ and their fluctuations $\boldsymbol{\beta^{-1}}$ remains Gaussian, with unit variance for $\boldsymbol{\beta^{-1}}$ and with variance for the features $\mathbf{x}$ inversely proportional to temperature—consistent with the fluctuation theorem. However, now one has the capability in TM to generate samples from the Gaussian prior at \textit{any} bath temperature $\beta$, including those at which no data was collected. By using the reverse SDE Eq. \ref{eq:tm_backward} data from the prior can then be mapped to data from the canonical ensemble at any temperature of interest irrespective of whether simulations were performed at that temperature or not.

\subsection{Exponentially Tilted Thermodynamic Maps (expTM)}
\label{sec:expTM}
As discussed in Sec.\ref{sec:ThermodynamicMaps}, the original TM framework\cite{herron2024inferring} established the connection between diffusion models and the fluctuation theorem, demonstrating how thermal fluctuations in the canonical ensemble correspond to variance in temperature. While useful, the original TM framework lacks the flexibility to capture fluctuations in different thermodynamic variables such as pressure and chemical potential, as well as the ability to capture fluctuations in more than one thermodynamic variables at the same time. To address this limitation, we extend the TM framework by introducing an exponentially tilted prior distribution, enabling its applicability across different thermodynamic ensembles. This generalization allows us to explore ensembles beyond the canonical system, including the isothermal-isobaric and Grand Canonical ensembles. 

\subsubsection{Isothermal-Isobaric thermodynamic maps}
\label{sec:expTM_NPT}
In the isothermal-isobaric (NPT) ensemble, the system is in thermal and mechanical equilibrium with a bath at temperature $T$ and pressure P, allowing fluctuations in both energy and volume.~\cite{Chandler1985Introduction} For such an ensemble, the probability for any microstate is given by: 
\begin{equation}
    p(x) = \frac{1}{Z} e^{-\beta E(x)} e^{-\lambda V(x)},
\end{equation}

\noindent where $\lambda = \beta P$, $Z$ represents the partition function and $V(x)$ is volume as a function of the microstate $x$. To adapt variations in data resulting from changes in both volume and energy into a diffusion framework, the key is to map the energy function $ E(x) $ and the volume function $ V(x) $ to a prior so that both can be controlled simultaneously. Once again we assume a simple, tractable prior with, but now an additional linear term for the volume. We note that in a different context, a similar idea was also explored in Refs. \onlinecite{vanleeuwen2023boltzmann, Schebek2024Efficient}. 
To incorporate the linear term, we introduce a parameter $\alpha =- P$  to simplify the expressions, and in the same spirit as Eq. \ref{eq:nvt_ensemble2}, express the prior as $p(x) = \frac{1}{Z} e^{-\frac{\beta}{2} x^2} e^{- \beta \alpha x}$. After completing the square in the exponent, this can be written as:
\begin{equation}
    p(x) =  \frac{1}{Z_{II}} e^{-\frac{\beta}{2}\left( x + \alpha \right)^2}
\label{eq:NPT probabilbity}
\end{equation}

\noindent where $Z_{II} = \int e^{-\frac{\beta}{2} \left( x + \alpha \right)^2} dx =\sqrt{\frac{2\pi}{\beta}}$  represents the partition function of the NPT ensemble.

Here $x$ denotes a single component feature and its fluctuation. In a  $D$-dimensional feature space,  $\mathbf{x} =  \{ x_1, \ldots, x_D\}$, where in the prior we assume that each component $x_i$  fluctuates independently of the others. Hence, the full prior factorizes as, 

\begin{equation}
    p(\mathbf{x}) \;=\; \prod_{i=1}^D p(x_i) 
    \;=\; \prod_{i=1}^D \frac{1}{Z_{II}} 
      \exp\!\Bigl[-\frac{\beta}{2} \,\bigl(x_i + \alpha\bigr)^2\Bigr].
\label{eq:NPT probability vector1}
\end{equation}

\noindent  Equivalently written as,

\begin{equation}
p(\mathbf{x}) \;=\; \frac{1}{Z_{II}^D} 
  \exp\!\Bigl[-\frac{\beta}{2} \,\sum_{i=1}^D \bigl(x_i + \alpha\bigr)^2\Bigr],
\label{eq:NPT probability vector2}
\end{equation}

\noindent where $\bigl(Z_{II}\bigr)^D = \Bigl(\sqrt{\tfrac{2\pi}{\beta}}\Bigr)^D$ represents the partition function for NPT ensemble. From Eqs.~\ref{eq:NPT probabilbity}-\ref{eq:NPT probability vector2}, one obtains $\langle \mathbf{x}\rangle = -\alpha$ and $\langle \mathbf{x}^2\rangle = \beta^{-1}$, which reflects exponentially tilted Gaussian prior distribution such that mean and variance of the prior correspond respectively to the pressure and temperature in the training data. Simultaneously, Eqs.~\ref{eq:NPT probabilbity}-\ref{eq:NPT probability vector2} indicate that the mean of the prior distribution for any given feature is shifted by $\alpha = - P$, showing how the pressure variations in the underlying ensemble from which training data was collected can be mapped to the prior. Thus, incorporating the pressure term into the energy allows the prior distribution to represent the isothermal-isobaric ensemble within the diffusion model framework.

\subsubsection{Grand canonical thermodynamic maps}
\label{sec:expTM_muVT}

In the Grand Canonical ($\mu$VT) ensemble, the system can exchange both energy and particles with a reservoir, allowing both the energy $E$ and number of particles $N$ to fluctuate.~\cite{Chandler1985Introduction}  The probability of finding the system in a specific microstate $ x $ is given by:
\begin{equation}
    p(x) = \frac{1}{Z} e^{-\beta E(x)} e^{\lambda N(x)}
\end{equation}

\noindent where $Z$ represents the partition function, $N(x)$ is the number of particles as a function of $x$,  $\lambda = \beta \mu$ with $\mu$ denoting chemical potential.

To map this to a prior distribution in feature space $\mathbf{x}$, we again maintain the harmonic potential energy to be $E(x) =\frac{x^2}{2}$ which is the same as in TM and for the NPT ensemble.  We define the particle density as $\rho(x) = \frac{N(x)}{V(x)}$. To incorporate variations corresponding to $N$ in the prior, we note that in the thermodynamic limit $N$ varies linearly with $V$, and thus we use the same linear treatment for $N$ as we did for $V$ in Sec.~\ref{sec:expTM_NPT} for the Isothermal-Isobaric thermodynamic maps. Once again we introduce a parameter $\alpha = \mu$ and express the prior as
$p(x) = \frac{1}{Z} e^{-\frac{\beta}{2} x^2} e^{\beta\mu x}$. Rearranging this and generalizing to $D$-dimensional features $\mathbf{x} =  \{ x_1, \ldots, x_D\}$ gives:
\begin{equation}
p(\mathbf{x}) 
\;=\; \frac{1}{(Z_{III})^D} 
  \exp\!\Bigl[-\frac{\beta}{2} \,\sum_{i=1}^D \bigl(x_i - \alpha\bigr)^2\Bigr].   
\label{eq:muPT probability vector}
\end{equation}

\noindent where $(Z_{III})^D = \int e^{-\frac{\beta}{2} (x - \alpha)^2} dx$ =$\Bigl(\sqrt{\tfrac{2\pi}{\beta}}\Bigr)^D$ represents the partition function for the $\mu$VT ensemble. We have again assumed the fluctuations in each coordinate are independent.
From Eq.\ref{eq:muPT probability vector}, we can extract $\langle x\rangle = \alpha$ and $\langle x^2\rangle = \beta^{-1}$, which reflects Gaussian exponential prior distribution such that mean and variance of the prior correspond respectively to the chemical potential and temperature in the training data. Simultaneously, Eq.\ref{eq:muPT probability vector}, incorporating the chemical potential term into the energy shifts the mean of the distribution by $\alpha =\mu$ allows the prior distribution in the diffusion model to represent the Grand Canonical ensemble.

\subsubsection{Exponentially Tilted -TM }
In the previous two Sec.\ref{sec:expTM_NPT} and \ref{sec:expTM_muVT}, we demonstrated how exponential tilting enables effective mapping onto different ensembles, extending the conventional TM framework. In the canonical ensemble, TM employs a Gaussian prior that encodes temperature fluctuations, which was limited to temperature variables. Here to summarize so far, we have introduced an exponentially tilted Gaussian prior, replacing the canonical prior $ p(x) = \frac{1}{Z} e^{-\beta x^2} $ with $ p(x) = \frac{1}{Z} e^{-\beta x^2} e^{\beta\alpha x} $, where $\alpha$ represents thermodynamic parameters such as $\alpha = -P$ or $\alpha = \mu$. This extension expTM enables the TM approach to capture fluctuations in NPT and $\mu$VT ensembles.

As demonstrated in Sec.~\ref{sec:Diffusion_models_exponentially_tilting}, the exponential tilting Gaussian can nicely merge into the OU process where the drift enforces the tilted Gaussian as its stationary distribution. Similarly, the original TM framework can be extended by incorporating the control parameter $\alpha$ while preserving the structure of the forward and reverse diffusion processes. To simplify the notation, we let $\boldsymbol{\eta}\equiv \{{\boldsymbol{\alpha},\boldsymbol\beta^{-1}}\}$. Here, the vectors $\boldsymbol{\beta^{-1}}$ and $\boldsymbol{\alpha}$ with same dimensionality as $\mathbf{x}$ quantifiy component-wise fluctuations in the feature space $\mathbf{x}$, as a function of the bath temperature and the bath pressure/chemical potential respectively. Then the forward process is written as: 
\begin{equation}
\begin{pmatrix}
\mathrm{d}\mathbf{x}\\
\mathrm{d}\boldsymbol{\eta}
\end{pmatrix}
=
-g(t)
\begin{pmatrix}
\mathbf{x}-\boldsymbol{\alpha_0}\\
\boldsymbol{\eta}
\end{pmatrix}
\mathrm{d}t
+ \sqrt{2 g(t)} 
\begin{pmatrix}
\sqrt{\boldsymbol{\beta_0^{-1}}} \\
\mathbf{1}
\end{pmatrix}
\mathrm{d}\mathbf{B}_t,
\label{eq:tilting_forward_force}
\end{equation}

\noindent where $\alpha_0$ represents the bath pressure or the chemical potential of the training input that shifts the feature-space mean, and $\beta_0^{-1}$ denotes the bath temperature analogous to the original TM. With the introduction of an additional exponential titling factor, the model is now parameterized in the joint space of $(\mathbf{x},\boldsymbol{\eta} ) \in \mathbb{R}^{3D}$. The reverse SDE converts samples from the prior distribution to match the target distribution. By incorporating an exponential tilting factor, we unify the TM framework across multiple thermodynamic ensembles without sacrificing tractability or simplicity properties of the Gaussian distribution. Following the original TM, the reverse SDE is written as
\begin{equation}
\begin{aligned}
\begin{pmatrix}
\mathrm{d}\mathbf{x}\\
\mathrm{d}\boldsymbol{\eta}\\
\end{pmatrix}
=& 
-g(\tau)\left[
\begin{pmatrix}
\mathbf{x}-\boldsymbol{\alpha_0}\\\boldsymbol{\eta}\\
\end{pmatrix}
+ \mathrm{s}_\theta(\mathbf{x},\boldsymbol{\eta}, \tau)
\right]
\mathrm{d}\tau \\
& + \sqrt{2 g(\tau)} 
\begin{pmatrix}
\sqrt{\boldsymbol{\beta_0^{-1}}} \\
\mathbf{1}\\
\end{pmatrix}
\mathrm{d}\mathbf{B}_\tau,
 \label{eq:tilting_backward}
\end{aligned}
\end{equation}

\noindent which is analogous to Eq.~\ref{eq:tm_backward} but now includes a new variable $\boldsymbol{\eta}$.
In this formulation, $\boldsymbol{\eta}$ encapsulates both the inverse temperature $\boldsymbol{\beta}^{-1}$ and the control parameter $\boldsymbol{\alpha}$ (e.g., pressure or chemical potential). 
The key difference compared to the original TM framework is that the inclusion of an new thermodynamic variable $\boldsymbol{\alpha}$ introduces an additional fluctuation variable that acts on the mean of the prior distribution—aligning it with the underlying bath pressure or chemical potential—while the inverse temperature $\boldsymbol{\beta}^{-1}$ continues to affect the variance of the Gaussian prior, ensuring consistency with the bath temperature.
These fluctuations are essential, as they encode the system's response to variations in thermodynamic conditions, thereby enabling the generation of samples at the desired global (bath) condition. This additional degree of freedom makes the prior more expressive and adjusts the feature space not just in response to temperature but also to additional thermodynamic control variable.

Solving Eqs.\ref{eq:tilting_forward_force} and \ref{eq:tilting_backward} allows us to vary not only temperature but also pressure or chemical potential using training data from two distinct thermodynamic conditions. 
The overall architecture is summarized in Fig.~\ref{fig:Model architecture} and as shown in Fig.~\ref{fig:Model architecture}(a), the model leverages two datasets that contain complementary information on mean features. Our training methodology extends the original TM by incorporating an additional control parameter that maps the mean to either chemical potential or pressure. Once training is complete, the learned representation is mapped onto a tilted Gaussian prior distribution, as described in the previous section. The reverse SDE is solved using the new prior to infer the target distributions, enabling the generation of new samples at a desired thermodynamic condition. We applied our model to two applications to showcase its performance, and the results are presented in the subsequent section.

\section{Results}
\label{sec:Results}

\begin{figure*}[htbp]
\captionsetup{skip=6pt}
\includegraphics[width=\textwidth]{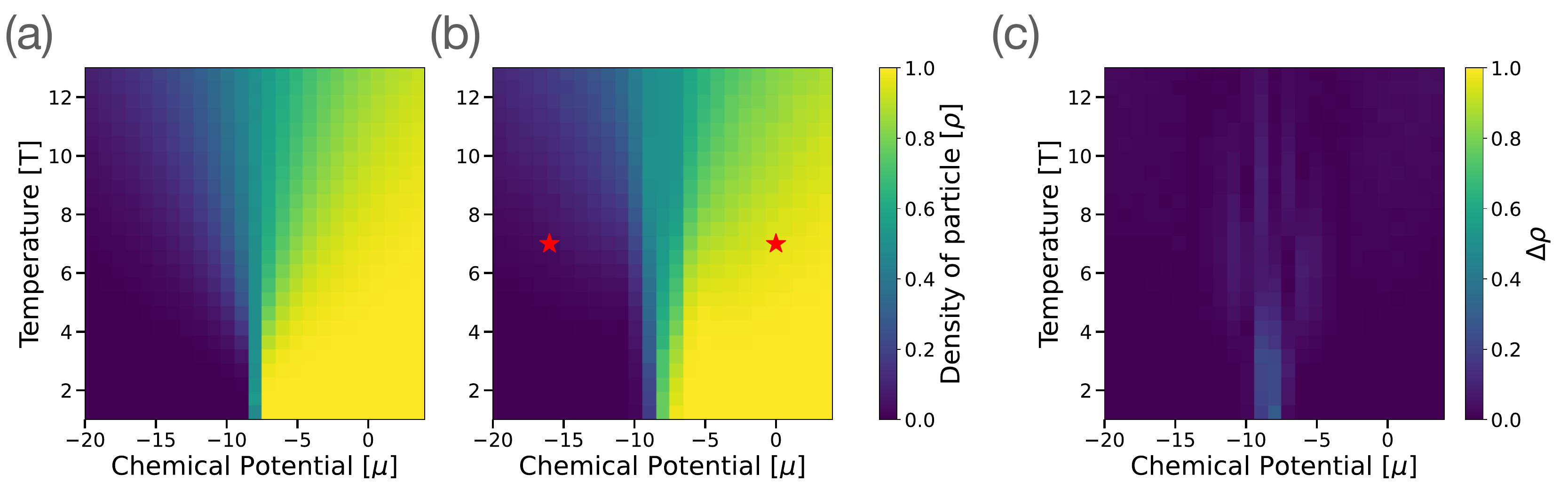}
    \caption{Exponentially Tilted (expTM) applied to simulate Grand canonical ensemble application for lattice gas model. 
    (a) Density of particles from benchmark Monte Carlo (MC) simulations at a wide grid of chemical potentials $\mu \in [-20,4]$ and temperatures $T \in [1,13]$, both sampled in steps of 0.5, yielding 625 grid points. 
    (b) Density of particles generated by the expTM approach. Only two sets of observations from the benchmark MC were used for training expTM, indicated as red stars at $\mu =-16, 0$ and $T = 7$. 
    (c)  Absolute difference in the density of particles ($\Delta \rho = |\rho_\text{MC}-\rho_\text{expTM}|$) between benchmark MC simulation and expTM generated results, indicating the high accuracy achieved by expTM.}
    \label{fig:GC ising model} 
\end{figure*}

\subsection{Grand Canonical Ensemble Application: Lattice Gas model}
\label{sec:lattice-gas}

In the previous work, TM demonstrated application using the simple Ising model, commonly used to describe phase transitions and critical behavior. In particular, we effectively reproduced the second-order phase transition of the heat capacity $C_v$ and extracted the critical temperature $T_c$, with training data only at two temperatures in the two phases far from the critical point. In the same spirit, here we first extend our work to a lattice gas model within the Grand Canonical (GC) ensemble. Before introducing the lattice gas model, we first revisit the Ising model.

In the standard Ising model, the energy function arises from spin-spin interactions, and with the presence of an  external magnetic field $H$, the Hamiltonian can be written as: 
\begin{equation}
    E_\text{spin} = -J \sum_{<i,j>} s_i \,s_j -H \sum_{<i,j>} s_i.
\end{equation}

\noindent Here, $ J $ denotes the coupling constant, and $ s_i \in \{+1, -1\} $ represents the spin variable at site $ i $. This system is well-suited to be simulated using the canonical ensemble system, which assumes a fixed number of particles. 

In certain cases, however, the system requires an exchange of both energy and particles with a reservoir. In such instances, the GC ensemble provides an appropriate framework, allowing the number of particles to fluctuate.\cite{lee1952statistical} Instead of spin–spin interactions, the GC ensemble describes whether lattice sites are occupied or unoccupied. This formulation, which characterizes interactions among neighboring pairs of particles, is referred to as a lattice gas. The transformation from an Ising model in the canonical ensemble to a particle-based lattice-gas model in the GC ensemble follows:
\begin{equation}
n_i = \frac{s_i + 1}{2}, \quad J = \frac{\epsilon}{4}, \quad H = \mu + 2JZ ,
\label{eq:GC mapping}
\end{equation}

\noindent where, $n_i \in \{0, 1\} $ represents the number of particles at site $i$, $\epsilon$ represents the pairwise nearest-neighbor interaction energy, and  $H$ is mapped to a chemical potential $\mu$ plus a coordination number $Z$. Applying Eq.~\ref{eq:GC mapping} mapping relation yields the Hamiltonian of the lattice gas (LG) energy function, expressed as:

\begin{equation}
    E_\text{LG} = -\epsilon \sum_{<i,j>} n_i \,n_j -\mu\sum_{<i,j>} n_i.
\end{equation}

We first performed a Monte Carlo Metropolis-Hastings (MC) simulation, which provides both our benchmark and the training dataset. While the original Ising model as studied with TM in Ref. \onlinecite{herron2024inferring} was controlled solely by temperature, the lattice gas model now depends on two parameters, namely the temperature and chemical potential. Varying either of the parameters changes the lattice occupancy and hence the particle density.

Figure~\ref{fig:GC ising model}(a) shows our benchmark MC calculations across different temperatures ($T$) and chemical potential ($\mu$) with color indicating particle density.
The result indicates that at low temperatures the particle density changes sharply near a critical chemical potential, whereas at high temperatures, the density varies more smoothly.  In analogy to the Ising model’s critical temperature, the lattice gas exhibits a critical chemical potential  ($\mu_c$). This critical point is observed when the $H = \mu + 2JZ = 0$, corresponding to $\mu_c = -8$. The MC simulations effectively capture both the sharp and the more gradual transitions in density.

Building on this framework, we applied our expTM approach an extension of TM that incorporates exponential tilting to the lattice gas model (Sec.~\ref{sec:expTM}). As we described in Sec.~\ref{sec:expTM}, a key advancement of expTM over the original TM is the ability to generate samples across two control parameters, i.e., temperature and chemical potential, rather than just temperature. We chose our feature space to be the lattice gas input variables, which we simulated as a two-dimensional $ N \times N $ lattice ($ N = 20 $), where each lattice site $ i $ is either occupied ($ n_i = 1 $) or unoccupied ($ n_i = 0 $). Each specific lattice site occupation represents a distinct microstate, denoted by the feature vector $\mathbf{x} = { n_1, n_2, \dots, n_{20\times20} } \in \mathbb{R}^{20\times20}$. From these lattice configurations, we construct a training dataset composed of sets ${\mathbf{x}, \boldsymbol{\mu}, \boldsymbol{\beta}}$. This dataset is then fed into the expTM framework and enables the model to learn how occupancy patterns fluctuate as $\boldsymbol{\mu}$ and $\boldsymbol{\beta}$ vary. The model captures phase transitions through changes in site occupancy as temperature and chemical potential conditions vary.

To implement the expTM method, we trained the system using two data points at a single temperature $T = 7$ but at different chemical potential values, specifically $\mu = -16$ and $\mu = 0$, marked as red stars in Fig.~\ref{fig:GC ising model}(b). The chemical potentials were deliberately chosen to be symmetric around the critical chemical potential. However, similar to results shown in Ref. \onlinecite{herron2024inferring} we expect asymmetric data points would also be effective, provided they are positioned on either side of the critical chemical potential. During the training, the temperature variable is mapped to the fluctuations of variance in the prior distributions, while the chemical potential enters as a shift in the mean of the prior.  
Despite being trained only at one temperature $T=7$, expTM can explore any combination of $(T, \mu)$, demonstrating the ability to handle two control parameters.

Figure~\ref{fig:GC ising model}(b) demonstrates expTM’s ability to generate lattice gas configurations across a broad range of $(T, \mu)$ conditions, accurately reproducing the critical boundary at $\mu_c=-8$. To quantify numerical accuracy, we compared expTM-generated particle densities with those from MC (Fig.\ref{fig:GC ising model}(a) vs. (b)) and computed ($\Delta \rho = |\rho_\text{MC}-\rho_\text{expTM}|$). As shown in Fig. \ref{fig:GC ising model}(c), the density difference remains within $\rho < |0.05|$ except in the immediate vicinity of $\mu_c=-8$. These results confirm that expTM reliably captures the phase transition in the lattice gas model, showcasing its key advantage: the ability to reproduce thermodynamically relevant samples by varying two control parameters simultaneously, thus extending its reach beyond the canonical ensemble.

\subsection{Isothermal-Isobaric Ensemble Application: CO$_2$ Phase transition prediction}
\label{sec:co2}

\begin{figure*}[htbp]
\captionsetup{skip=6pt}
\includegraphics[width=\textwidth]{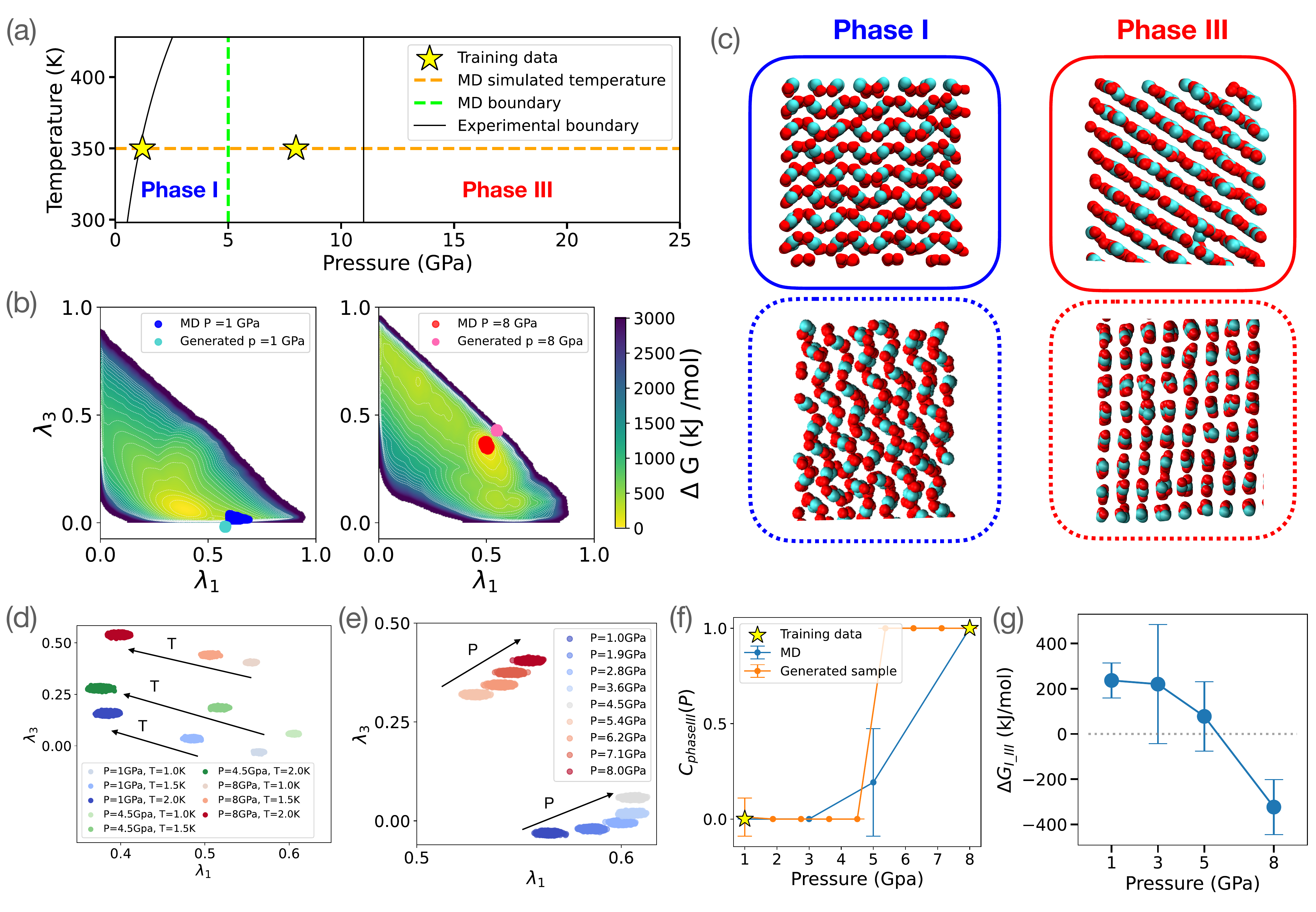}
    \caption{ expTM applied to simulate the Isothermal-Isobaric ensemble for predicting CO$_2$ phase transitions. 
    (a) Phase diagram of CO$_2$ in temperature range of 300–400 K and pressure range of 0–25 GPa. The solid line represents the experimental phase boundary as defined in Ref.~\cite{datchi2009structure,iota2001phase,santoro2006high}. The green dotted line corresponds to the phase boundary determined by molecular dynamics (MD) simulations with enhanced sampling methods.~\cite{Ilaria2017CO2} The orange line indicates the temperature used in our simulations. The two yellow stars denote the training data points used to train expTM. 
    (b) Free energy profiles at two pressures, P=1 and P=8 GPa. In each plot, the training data is labeled as ``MD" and highlighted in blue and red for P=1 and P=8 respectively. Generated data from expTM is labeled as ``generated" and visualized in sky blue and pink for P=1 and P=8 respectively. 
    (c) CO$_2$ crystal structures observed in simulations. The first row (solid squares)  shows the reference configurations from MD simulations. The second (dotted squares) row depicts the CO$_2$ phases extracted after back-mapping expTM generated features to the MD simulation trajectories. Details on the back mapping are provided in the SI. (d) Phase predictions from expTM were obtained over a range of TM-based temperatures—where each unit of 1 K corresponds to 350 K and increases linearly—and at three different pressures. 
    (e)  Phase predictions from expTM over at fixed temperature (350 K) and at three different pressures. Phase III regions are shown at the top, and Phase I regions at the bottom. 
    (f) Closeness to Phase III ($C_\text{phasesIII}(P)$), quantified as a ratio where 1 indicates complete alignment with Phase III and 0 corresponds to Phase I. Yellow stars mark the training data, while other markers show expTM results. The phase boundary for MD is determined as shown in Table ~\ref{tab:phaseboundary}. Specifically, we use a $\lambda_1$  threshold of 0.58, which lowers the minimum range for Phase I and extends the range for Phase III. This adjustment accommodates the wide spread of the generated data and ensures a clear distinction between the phase states, reducing numerical inconsistencies.    (g) Free energy differences between Phase I and III  obtained from MD simulations. A positive $\Delta G_{I\_III}$ suggests that phase I is more stable than phase III and vice versa.}
    \label{fig:co2_combine} 
\end{figure*}

As a second application of expTM, we study the different structural arrangements of CO$_2$ under various pressures and temperatures. CO$_2$ is of broad interest due to its diverse molecular structural arrangements under different conditions of pressure and temperature.\cite{Ilaria2017CO2} 
Despite its importance, the complete phase diagram remains elusive, particularly under the extreme pressures and temperatures regime where experimental measurements become very challenging. From a computational standpoint, accurately capturing relevant phases of CO$_2$  demands extensive sampling and precise calculations, especially when dealing with extreme conditions or supercritical regimes.\cite{Ilaria2017CO2} In this part of the study, we demonstrate how expTM can successfully deal with aspects of this challenge. We show the capability of expTM to reproduce important phase boundaries of CO$_2$ by using a limited set of training data.

To examine the pressure-dependent states of solid CO$_2$, we investigate its polymorphic transitions, which are governed by external pressure and temperature. The precise delineation of the phase boundaries remains a topic of ongoing debate; however, experimental studies consistently identify Phase I and Phase III as distinct states.~\cite{datchi2009structure,iota2001phase,santoro2006high} Notably, the phase boundary separating these two phases has been reported at approximately 11.8 GPa~\cite{datchi2009structure}. These phases are primarily distinguished by their molecular orientations and structural symmetries. Phase I exhibits a cubic $Pa\overline{3}$ symmetry, while Phase III adopts an orthorhombic Cmca structure, transitioning through a coordinated molecular rearrangement.~\cite{santoro2006high,aoki1994crystal,bonev2003high}  In Fig.~\ref{fig:co2_combine}(a), we report the experimentally well-established narrow phase boundary range T = 300–400 K and P=1–20 GPa represented as a solid black line.\cite{datchi2009structure}

Although computationally challenging, the transition between Phase I and Phase III has been successfully demonstrated by Ref.\onlinecite{Ilaria2017CO2} using well-tempered metadynamics.\cite{Barducci2008well, Barducci2011metadynamics} Building on these insights, we reproduced these results and extended the analysis by performing benchmark simulations at a temperature of 350 K and pressures of 1, 3, 5, and 8 GPa. Details of the MD simulation setup, including the choice of collective variables (CVs) are provided in the Supporting Information (SI).

Before discussing the details of expTM for this system, we first summarize the key findings and clarify important parameters discussed in Ref.~\onlinecite{Ilaria2017CO2} and other literature.
Unlike experiments, which typically observe only one stable phase at each pressure, enhanced sampling MD simulation captures multiple states—including Phases I, III, and an amorphous liquid phase. However, to accurately evaluate the relative stability of these phases during simulations, it is necessary to reweight the biased ensembles appropriately.\cite{Ilaria2017CO2} Under low-pressure conditions (P = 1 GPa), Phase I is most stable, whereas at high pressure (P = 8 GPa), Phase III becomes dominant. To distinguish these phases, the CVs $\lambda_1$ and $\lambda_3$ serve as robust descriptors.  Here, $\lambda_1$ and $\lambda_3$ quantify the average similarity of each CO$_2$ molecule to Phases I and III, respectively, with values constrained between 0 and 1 (see the SI for details). Importantly, both descriptors must be considered together as a pair to differentiate between phases. Each phase is associated with specific ranges of these CVs, as detailed in Table\ref{tab:phaseboundary}. Generally, high $\lambda_1$ and low $\lambda_3$ indicate Phase I, while the opposite suggests Phase III. Once these characteristic ranges are established, one can estimate the free-energy difference between Phases I and III ($\Delta G{I\text{-}III}$), as illustrated in Fig.~S1 of the SI.

\begin{table}[ht!]
\begin{tabularx}{0.5\textwidth}{XXX}
\hline
    Phases & $\lambda_1$ range  & $\lambda_3$  range \\
\hline
Phase I &  $\lambda_1>0.6$  &  $\lambda_3<0.25$ \\
Phase III & $\lambda_1<0.55$  &  $\lambda_3>0.28$ \\

\hline
\end{tabularx}%
\caption{\label{tab:phaseboundary}%
The phase boundaries of phase I and III are defined in terms of $\lambda_1$ and $\lambda_3$.}
\end{table}

Following the idea discussed in Sec.~\ref{sec:expTM}, we encode the two CVs $\lambda_1$ and $\lambda_3$ atomwise into a feature matrix $\mathbf{x}$. Each atom in the system is assigned its own values, $(\lambda_1)_i$ and $(\lambda_3)_i$, which quantify its similarity to Phases I and III, respectively. These atom-wise values provide local structural details, and averaging them over all atoms yields a global measure of the phase behavior. In our simulation of a 256-atom system, the complete set of $\lambda_1$ and $\lambda_3$ values forms two feature vectors:  
\begin{itemize}
\item $\mathbf{x}_1 = \{(\lambda_1)_1, (\lambda_1)_2, \ldots, (\lambda_1)_{256}\} \in \mathbb{R}^{256}$, and
\item
 $\mathbf{x}_3 = \{(\lambda_3)_1, (\lambda_3)_2, \ldots, (\lambda_3)_{256}\} \in \mathbb{R}^{256}$.
\end{itemize} 
Here, $\mathbf{x}_1$ and $\mathbf{x}_3$ capture the detailed local structural descriptors for each CO$_2$ molecule in the system. Each feature matrix is coupled to the corresponding bath temperature ($\beta_0$) and bath pressure ($P_0$) at which it was sampled. We then treat $(\mathbf{x}_1, \mathbf{x}_3, \mathbf{P}, \boldsymbol{\beta})$ as the training data for expTM. Once training is complete, we can generate new $\mathbf{x_1}$ and $\mathbf{x}_3$ matrices at any desired temperature and pressure, recovering the atom-wise $\lambda_1$ and $\lambda_3$ values. This approach allows us to predict the behavior of the system under conditions not present in the initial dataset.

To prepare our training dataset, we first select configurations from MD simulations at $ P = 1 $ GPa and $ P = 8$ GPa. For each pressure, $10^5$ sample configurations are selected according to the Boltzmann distribution based on the free energy surface (Fig.~S1). 
At each pressure, the free energy difference between the two phases is significant (Fig.~\ref{fig:co2_combine}[g]) such that selecting configurations for training dataset near the energy minimum ensures only one phase is sampled for each pressure (Fig.~S2). The selected pressure and temperature variables are marked by yellow stars in the phase diagram in Fig.~\ref{fig:co2_combine}[a]. The selection of features $\mathbf{x_1}$ and $\mathbf{x}_3$, defined by $\lambda_1$ and $\lambda_3$, is shown in Fig.\ref{fig:co2_combine}[b] as red and blue dots, respectively. 
To illustrate the distinct molecular arrangements in Phases I and III, we present representative conformations in Fig.~\ref{fig:co2_combine}[c].

The expTM model is trained on the dataset with two primary objectives:  (1) to recover Phases I and III across varying pressures and temperatures and (2) to delineate the phase transition boundary, thereby identifying intermediate states. 

The results addressing our first objective are shown in Fig.~\ref{fig:co2_combine} [d] and [e]. Using our training dataset, new data has been generated based across a range of pressures and temperatures. Fig.~\ref{fig:co2_combine}[d] illustrates the generated data at three different temperatures ($T = 1, 3, 5\, \mathrm{K}$) under varying pressures ($P = 1, 4.5, 8\, \mathrm{GPa}$). Additionally, Fig.~\ref{fig:co2_combine}[e] depicts results over a broader pressure range ($P = 1$ to $8\, \mathrm{GPa}$). Notably, at $P = 4.5\, \mathrm{GPa}$, the system is identified as being situated between Phase I and Phase III.  To validate these findings, the closeness to Phase III was evaluated using a metric ranging from 0 to 1. A closeness value near 1 indicates proximity to Phase III, whereas a value near 0 signifies a state closer to Phase I. This metric was computed based on the given $\lambda_1$ and $\lambda_3$ ranges to assess how closely the data aligns with the defined phase boundaries. 

Our second objective was achieved by generating samples across the pressure range, revealing that the intermediate state occurs between approximately $P = 4.5 - 5.4\, \mathrm{GPa}$, where a sharp transition is observed in the data as shown in Figs.\ref{fig:co2_combine}(d) and (f). This result is consistent with expectations on the simulated data, which span
$P = 1 - 8\, \mathrm{GPa}$, and effectively capture the symmetry at the midpoint. Moreover, despite the discrete pressure values, the expTM method successfully identified sharp peaks and transitions between adjacent pressure states.

\section{Conclusion}
\label{sec:conclusion}
In conclusion, here we have presented Exponentially Tilted thermodynamic map (expTM), a Generative AI approach tightly integrated with principles of statistical physics, that makes the best of limited training data to generate more samples under differing environmental conditions with the correct equilibrium distribution.
expTM is an extension of the original TM framework introduced in Ref.~\onlinecite{herron2024inferring}. While the TM approach allows generating samples at different temperatures following the canonical ensemble, the expTM framework generalizes this framework to different ensembles including Grand Canonical and Isothermal-Isobaric ensembles.
This advancement provides the flexibility to generate new samples conditioned not only on temperature but also on additional thermodynamic variables of interest, such as pressure or chemical potential.
The enhanced expressiveness of the expTM is achieved by integrating an exponential tilting factor into the Gaussian prior distribution within the original score-based model, resulting in a more comprehensive representation of the prior distribution.
We validated the significance of the model through key applications: specifically, the lattice gas model under the Grand Canonical ensemble and CO$_2$ phase transitions in the isothermal-isobaric ensemble.

In the Grand Canonical ensemble, expTM was trained on just two data points which reproduced phase transitions in the lattice gas model on two control variables i.e., temperature and chemical potential. expTM accurately generated the density of particles with minimal deviation from MC simulations, except near critical points. In the isothermal-isobaric ensemble, expTM predicted phase transitions in solid CO$_2$ under varying pressure and temperature, identifying phase boundaries and intermediate states between Phase I and Phase III. This highlights its effectiveness in capturing structural arrangements in high- and low-pressure regions beyond MD simulation capabilities.

Beyond the two demonstrated applications, the expTM framework can show significant promise for systems that are challenging to model with conventional MC or MD methods. For instance, it may be applied to spin-glass models, crystal nucleation processes or intrinsically disordered proteins and RNA that show significant dependence on environmental factors such as temperature, pressure, and pH. We are pursuing these and other applications. Finally, while our study focuses on an exponentially tilted diffusion model to capture phase transitions across different thermodynamic ensembles, recent work has shown connections between diffusion models and stochastic localization~\cite{montanari2023sampling}, where the latter may serve as a closed-form method to exponential tilting within a continuous time-dependent framework~\cite{albergo2023stochastic}. In particular, Ref.\onlinecite{ghio2024sampling} showed that applying an exponential tilting can reproduce the behavior of spin glass systems~\cite{Edwards1975Theory}, indicating that exponential tilting could be a valuable tool for describing a system under various thermodynamic conditions. Although we did not take the norm of coordinate variable $x$ as proposed in Ref.\onlinecite{floto2023the}, incorporating the following refinement, along with the stochastic localization framework, has the potential to further enhance our model.   \newline

\textbf{Supporting information\newline }
Further details on the Lattice gas model setup, CO$_2$ simulation setup, and collective variables (CVs) can be found in the Supporting Information (SI). 
\newline

\textbf{Acknowledgements\newline }
This research was entirely supported by the US
Department of Energy, Office of Science, Basic
Energy Sciences, CPIMS Program, under Award
DE-SC0021009. We thank UMD HPC’s Zaratan and NSF ACCESS (project CHE180027P) for computational resources. P.T. is an investigator at the University of Maryland-Institute for Health Computing, which is supported by funding from Montgomery County, Maryland and The University of Maryland Strategic Partnership: MPowering the State, a formal collaboration between the University of Maryland, College Park, and the University of Maryland, Baltimore. We would like to thank Akashnathan Aranganathan for insightful discussions and Richard John for carefully reviewing the manuscript. 
 \newline

\textbf{Data availability statement\newline }
A Python implementation of the Grand Canonical lattice gas model will be released on GitHub prior to submission to a peer-reviewed journal: \href{https://github.com/Suemin-Lee/ExpTM_GC}{www.github.com/Suemin\-Lee/ExpTM\_GC} All associated files will also be provided on PLUMED-NEST.
\newline

\textbf{References}

\end{document}